# STUDENTS' DIFFICULTIES IN UNDERSTANDING THE BASIC PRINCIPLES OF RELATIVITY AFTER STANDARD INSTRUCTION

*Dimitrios Gousopoulos, Efstratios Kapotis and George Kalkanis*

University of Athens, Pedagogical Department P.E.,Science, Technology and Environment Laboratory

*Abstract:* Theory of Relativity (Special and General) is one of the most influential theories of the 20th century and has changed the way we view the world. It is part of many undergraduate curriculums and it is often suggested that it should be integrated into an upper secondary curriculum. Special Theory of Relativity combines time and space whereas General Theory of Relativity describes gravity as a geometric property of spacetime. As it describes abstract phenomena, students encounter several difficulties understanding its basic principles and consequences. In this paper, we present the research that we conducted in order to detect the aforementioned difficulties. This research constitutes a part of a more general study concerning the integration of Special and General Relativity into an upper secondary and undergraduate curriculum. The sample consisted of 45 non-major physics undergraduate students. The purpose of our study was to determine and categorize the difficulties students face when they study the principles of the Theory of Relativity and its consequences. The results of our research indicate that student face many obstacles when trying to interpret phenomena described by the Relativity and confuse Special and General Relativity principles. These results dictate us to create an educational approach that tackle the difficulties found.

## INTRODUCTION

Theory of Relativity is one of the most influential theories of the 20th century that change the way we view the physical world. It is part of many undergraduate curriculums and it is often suggested that it should be integrated into an upper secondary curriculum (Arriassecq & Greca, 2010; Villani & Arruda, 1998). The failure of experiments to detect any motion of ether led to Einstein's two basic postulates of Special Relativity and their consequences (relativity of simultaneity, time dilation and length contraction). While Special theory of Relativity is primarily concerned with inertial frames of reference, accelerated frames of reference and gravity are treated in General Relativity, as table 1 presents. Relativity has, also, many applications with GPS being one of the most known among them. As the phenomena Relativity describes can be characterized as abstract, researching students' difficulties in understanding these phenomena is of great interest for the educational community.

Table 1.
*The Theory of Relativity*

| Theory of Relativity | |
|---|---|
| Special Relativity | General Relativity |
| Basic Postulates | Basic Postulate |
| Invariance of Physical Laws<br>Invariance of the speed of light | Principle of Equivalence |
| Applications | Applications |
| Relativity of Simultaneity<br>Time Dilation<br>Length Contraction | Bending of light<br>Modification of time by gravitational fields |

## Reasons why we should teach relativity and conduct research on how best to include it in current curriculum

Relativity constitutes a revolutionary & influential Theory, it is part of many undergraduate curriculums and many researchers and educators suggest that it should be integrated into an upper secondary curriculum. It is part of our cultural heritage and describes as well as interprets abstract phenomena. Moreover, it provokes students' excitement and develops abstract way of thinking. So, researching students' difficulties in understanding these phenomena is of great interest for the educational community.

## LITERATURE REVIEW

### Defining the term "Difficulties in understanding"

According to Centeno (1988) difficulty is something that constrains students' understanding of a given subject. Some of students' ideas correctly interpret particular phenomena, but fail to do so in other phenomena (more general most of the time). According to Brousseau (1983) , difficulty is knowledge, not necessarily lack of it. In addition to this, students often ignore the existence of a wrong knowledge and as a result they find it difficult to replace it with the scientifically correct one.

### Educational Research in Relativity

There is a small number of studies concerning the difficulties students face when they study the Theory of Relativity (Pitts,Venville, Blair & Zadnik, 2014) the results of which are summarized in the following points:
a) Students find it difficult to define and thus describe a Frame of Reference and they believe in the existence of a privileged observer (Arriasecq I. & Greca M.I. 2010; Scherr et al 2001; Panse et al 1994 ;Ramadas et al 1996; Villani & Pacca, 1987).
b) Many students predict the progress of a physical phenomenon, without using the principle of relativity. Moreover, they cannot apply the invariance of the speed of light (Pietrocola & Zylbersztajn,1999; Scherr et al 2001;Dimitriadi & Halkia, 2012).
c) Students cannot perceive that two simultaneous events in a particular Frame of Reference are not necessarily simultaneous relative to another Inertial Frame of Reference. What is more, they hold the view that time is absolute and that both time dilation and length contraction constitute a distortion of the reality (Scherr 2007, Scherr et al 2001; Hewson 1982 , Posner et al 1982).

d) Students use Special Relativity postulates so as to interpret phenomena of General Relativity (Bandyopadhyay,A., Kumar, A., 2010).

## RESEARCH QUESTION

Aiming to contribute to the aforementioned literature we designed our research and developed our research question.

"Which are non-major physics students' difficulties in understanding the basic principles of the Theory of Relativity and its consequences after standard instruction?"

## METHOD

In order to measure the difficulties, we constructed a questionnaire that included open ended questions and was based on the existing literature. In addition to questionnaire (Figure 1), we conducted interviews so as to go deeper and get a better picture of students' difficulties. The research was conducted in two phases:
In the first phase we conducted a pilot study with 10 non major physics undergraduate students as our sample. Our tool was a questionnaire that included open ended questions and was revised by 2 experts in physics and 2 experts in educational research.

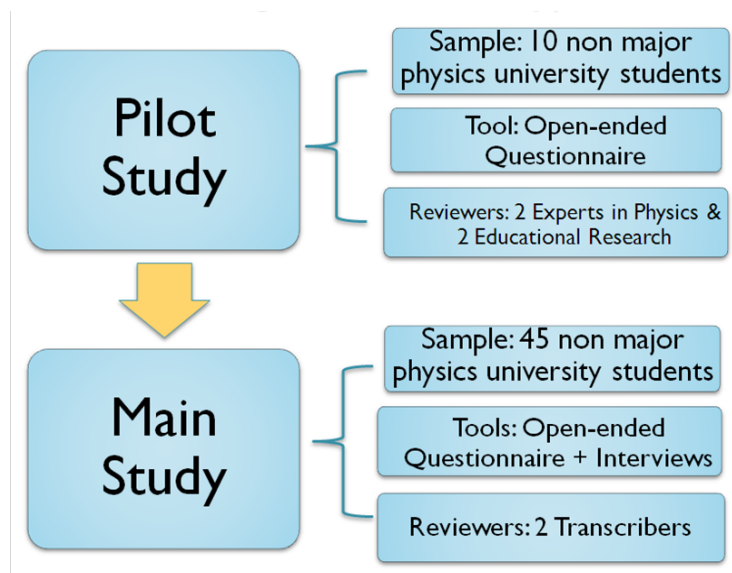

Figure 1. **Research Design – Qualitative Approach**

The next phase included the main study with a sample of 45 non major physics undergraduate students of the Pedagogical Department at the University of Athens. As our main tool we used the aforementioned questionnaire that that was amended so as to incorporate students' feedback from the pilot study. In addition to this questionnaire we conducted interviews with 10 students so as to get a deeper understanding of their difficulties and test our assumptions. Finally, aiming to assure the reliability of our study, we had 2 transcribers to analyze the data collected.
Students that comprised the sample of our main study attended a series of lessons, the structure of which is shown at Table 2.

Table 2.
*Regular Instruction Scheme*

| Lesson Unit | Instructional Time Interval |
| --- | --- |
| Kinematics & Dynamics Review | 2 x 45 min. |
| Inertial Frame of References | 2 x 45 min. |
| Basic Principles of the Special Theory of Relativity | 2 x 45 min. |
| Relativity of Simultaneity, Time Dilation, Length Contraction | 3 x 45 min. |
| Principle of Equivalence | 2 x 45 min. |
| Bending of light | 45 min. |
| Modification of time by gravitational fields | 45 min. |

The aforementioned Instructional Scheme is based on the newly introduced Greek Curriculum.
Our tools examined the following thematic units:
a) Principle of Relativity, b) Invariance of the speed of light, c) Relativity of Simultaneity, d) Time Dilation, e) Length Contraction, f) Principle of Equivalence,   h) Bending of light and i) Modification of time by gravitational fields.
After analyzing the data collected using qualitative approaches, we categorized our findings based on the above mentioned thematic units.

# RESULTS

## Einstein's Principle of Relativity - invariance of physical laws

Some students (40%) found it difficult to perceive the equivalence between motionless and uniform motion. They said that phenomena (either electromagnetic or mechanical) can progress differently for different observers. For instance, they believe that an object moving at a constant speed relative to an observer O can accelerate or decelerate relative to a different inertial observer O´.

## Invariance of the speed of light

Many students (66,7%) stated correctly the invariance of the speed of light, but they failed to apply it in problems in which the speed of light was demanded. Instead they used the Galilean velocity addition formula.

## Relativity of Simultaneity

Some students (24,4%) considered that two events that are simultaneous for an observer, must be simultaneous for any other inertial observer.
*"Since the 2 explosions occur simultaneously for observer A, they are simultaneous for any other inertial observer"*.

## Time Dilation

A common view among students (77,7%) was that time is absolute and unaffected by the inertial observer relative to whom it is measured.
*"Time is what it is, it can't change"*.

## Length Contraction

Students (15,5%) held the view that length is the object's inherent characteristic, thus it cannot change. They, also, related the object's length to its mass. They said that as mass cannot differentiate, nor can its dimensions.
*"Since mass don't change, length stays the same"*
Even though some students (71,1%) predicted correctly the length contraction, the reasons they projected were wrong. The contraction was attributed to the high speed.
*"The spaceship is moving so fast that it <u>seems</u> smaller"*
Finally, some students (55,5%) while predicting correctly the length contraction, they said that this contraction occurs in every direction of the object's motion
*"...the object's length has been shortened, as it moves at x axis; likewise, there would be a length contraction if the object moved at y axis"*

## Shift in the inertial observer

Students (44,4%) found it difficult to extract correct conclusions when the inertial observer, to whom a relativistic phenomenon takes place, changes. Many students replied correctly when the phenomenon they examined took place at a Frame of Reference that is motionless relative to the observer. Yet, when the phenomenon took place at a Frame of Reference that moved uniformly relative to the observer, wrong answers rose sharply

## Boundaries between classical Mechanics and Relativity

A prevailed view among students (73,3%) was that Relativity deals with phenomena that occur at relativistic speeds or due to large amounts of mass.
*"Theory of relativity concerns phenomena that take place only at speeds near the speed of light".*

## Principle of Equivalence

While applying correctly the Principle of Equivalence outside a gravitational field, students (64,4%) failed to do the same in areas inside a gravitational field.
*"If I accelerate a box towards the earth at amount equal to g, the people inside the box will feel doubled acceleration. We should accelerate the box by g at the opposite direction so that they can feel zero gravity".*

## Bending of light

Classical Physics was deeply rooted into students' mind (57,7%). There were cases where they changed a correct answer due to the aforementioned obstacle.
*"Light can't follow a curved path. The right lines are those that I haven't erased"*(Figure 2).
Moreover, bending of light was being attributed to the existence of an invisible mirror (51,1%).
*"There should be a mirror so as to make the light to curve"*

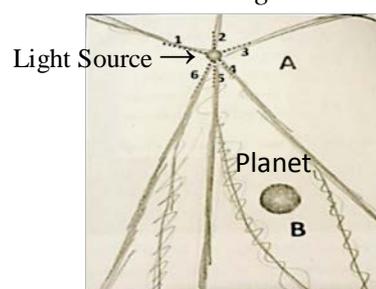

Figure 2. **Student's drawing concerning the bending of light**

**Modification of time by gravitational fields**

There was a confusion between General and Special Relativity regarding the rate of time flow inside a gravitational field (37,8%).
*"Since a satellite has a velocity at this height, there is a time dilation; is there any possibility to have a time modification due to both theories?"*

## DISCUSSION

Returning to our research question, we found that non-major physics students encounter various difficulties after regular instructional series. In general, most of our findings go along with the corresponding findings of the international literature. More specifically, students find it difficult to understand the invariance of physical laws confirming the results of Pietrocola & Zylbersztajn (1999). Moreover, students use the Galilean velocity addition formula so as to answer question regarding the invariance of the speed of light ignoring the invariance of the speed of light (Scherr et al 2001). As far as the consequences of Special Relativity are concerned, students face several difficulties understanding them. For instance, students stated that two events that are simultaneous for an observer must be simultaneous for any other inertial observer (Scherr et al 2001). Furthermore, students project the idea that time and space are absolute and unaffected by the inertial observer (Dimitriadi & Halkia, 2012; Posner et al 1982). An interesting aspect of our findings is that, the contraction was attributed to the high speed and even though some students predicted correctly the length contraction, they said that this contraction occurs in every direction of the object's motion. What is more, students found it difficult to extract correct conclusions when the inertial observer, to whom a relativistic phenomenon takes place, changes. Our research has, also, unveiled difficulties concerning the General Relativity. Students fail to apply correctly the principle of equivalence inside a gravitational field and attributed the bending of light to the existence of an invisible mirror. Moreover, among students there was confusion between General and Special Relativity regarding the rate of time flow inside a gravitational field. Finally, one finding that should be underlined is the view that Relativity deals with phenomena that occur at relativistic speeds or due to large amounts of mass constraining students understanding regarding the boundaries between classical Mechanics and Relativity. The aforementioned findings of our research combined with the relevant international literature should be taken into account when a country's educational committee intends to build a curriculum that includes topics from the Relativity. Acknowledging students' difficulties in understanding the basic elements of the theory of relativity, leads to the construction of an effective educational approach.
An interesting expand of our study would be the application of our tool to students coming from the upper secondary education, as well as from physics departments.
To conclude, an educational approach should be created aiming to cope with the difficulties found. Our proposal, which will be our next step, is an educational approach using interactive dynamic simulations in order to visualize relativistic phenomena that exists outside our everyday experience.